# Benchmarking Apache Spark and Hadoop MapReduce on Big Data Classification

TAHA TEKDOGAN

Department of Computer Engineering, Istanbul Technical University

Radar and Electronic Warfare Intelligence Systems Group, ASELSAN A.S.

ALI CAKMAK

Department of Computer Engineering, Istanbul Technical University

Most of the popular Big Data analytics tools evolved to adapt their working environment to extract valuable information from a vast amount of unstructured data. The ability of data mining techniques to filter this helpful information from Big Data led to the term 'Big Data Mining'. Shifting the scope of data from small-size, structured, and stable data to huge volume, unstructured, and quickly changing data brings many data management challenges. Different tools cope with these challenges in their own way due to their architectural limitations. There are numerous parameters to take into consideration when choosing the right data management framework based on the task at hand. In this paper, we present a comprehensive benchmark for two widely used Big Data analytics tools, namely Apache Spark and Hadoop MapReduce, on a common data mining task, i.e., classification. We employ several evaluation metrics to compare the performance of the benchmarked frameworks, such as execution time, accuracy, and scalability. These metrics are specialized to measure the performance for classification task. To the best of our knowledge, there is no previous study in the literature that employs all these metrics while taking into consideration task-specific concerns. We show that Spark is 5 times faster than MapReduce on training the model. Nevertheless, the performance of Spark degrades when the input workload gets larger. Scaling the environment by additional clusters significantly improves the performance of Spark. However, similar enhancement is not observed in Hadoop. Machine learning utility of MapReduce tend to have better accuracy scores than that of Spark, like around 2%-3%, even in small-size data sets.

**CCS CONCEPTS • Information systems → Data management systems**

**Additional Keywords and Phrases:** Big Data, Data Mining, Classification

## 1 INTRODUCTION

Nowadays, wide ranges of applications produce enormous amount of data that carry valuable information. Since there are no strict standards established for data; it may be in any size, arbitrary structure, and available for very restricted time to process. These characteristics led to the term "Big Data", which is too large and complex data to be processed by traditional data management systems. Challenges that emerged in the Big Data world can be grouped under "3V" s – volume, variety, and velocity [1]. Recent papers update this group by adding two extra dimensions, i.e. value and veracity [2]. Volume refers to the amount of data produced by applications. It may be very large such as Petabytes, especially when we need to process multimedia-related content. Variety stands for the format of the data generated by humans or machines, that may be structured or unstructured. An ideal data management system should be able to classify data into various categories. Velocity refers to the speed of the data being generated. In some cases, we may have limited time to process data and respond to the client, e.g., in a search engine scenario. In short, one should consider all these challenges to develop an ideal data management system that deal with Big Data.

On the other hand, there is an ever-increasing interest for data mining applications that are designed to extract helpful information to make high-impact decisions [3,4]. Many algorithms have been developed to explore unknown patterns, that implement varying techniques such as classification, clustering, prediction, and association rule mining. Earlier, these methods were designed to process small-to-medium scale structured data. Hence, these techniques have been improved so that they can handle Big Data. The term, "Big Data Mining", stands for the effort to adapt data mining operations to Big Data.

According to Jaseena et al. [5], Big Data Mining is the capability of extracting useful information from large datasets and streams of data, which was not possible before due to its volume, variety, and velocity. Since we intend to accommodate Data Mining operations on Big Data, we need to consider potential challenges of scaling the existing algorithms to operate on high volume data, handling the heterogeneity of data, and being able to catch up with the velocity of streaming data. There have been some frameworks to address these challenges and expose different approaches to successfully run data mining operations on Big Data, such as Apache Spark and Hadoop MapReduce. Even though these tools come handy, there are some pitfalls caused by their architectural limitations. One framework may be able to scale algorithms on huge amount of data but may have undesirably long running times, or vice versa. In this paper, we aim to identify strengths and pitfalls of Apache Spark and Hadoop MapReduce in big data mining illustrated through the classification task on various datasets.

Proposing appropriate methods to measure the performance of the underlying frameworks is crucial, since we want to identify job-specific features of the benchmarked tools. In this paper, since we focus on the classification problems, we define evaluation metrics in a specialized way accordingly. In particular, as metrics, we study *execution time*, i.e., duration of the training phase of the program; *accuracy*, i.e., predicted values' degree of closeness to actual values, and *scalability*.

To the best of our knowledge, there is no comprehensive previous research that studies the classification task on Big Data by taking task-specific metrics into consideration. Thus, we present a benchmark to understand the characteristics of these tools and compare them in terms of scalability, accuracy, and the duration of training phase. Unlike previous benchmark studies that generate the same input of varying sizes; we use four different datasets that present four varying workload sizes.

The remainder of this paper is structured as follows: Section 2 introduces the previous studies on benchmarking Spark and MapReduce, the gaps in the literature, and our contributions. Section 3 discusses the methods that we use in our comparative study such as frameworks and their utilities, implemented machine learning techniques, and our benchmarking methodology. In Section 4, we describe the datasets that we use in our experiments, define evaluation metrics to measure the performance of the benchmarked frameworks, and present the results of our study. Section 5 concludes the paper with possible future directions.

## 2 RELATED WORK

Shi et al. [6] provide a benchmark to evaluate major architectural components in Spark and MapReduce frameworks such as shuffle, execution model, and caching. They also provide two profiling tools: (i) 'Execution

Plan Visualization' to correlate task execution plan with resource utilization for both frameworks; and (ii) 'Fine-grained Time break-down' to visually present this correlation.

Samadi et al. [7] employ HiBench [8] to compare Hadoop and Spark in terms of performance based on criteria including execution time, speedup, and throughput. Their results show that Spark outperforms Hadoop, but with the trade-off of higher memory consumption. Another study of Samadi et al. [9] deploy Apache Spark and Hadoop MapReduce on virtual machines. The comparison criteria is the same as their previous study [7]. They test Word Count workload with varying sizes of data. Their results show that the performance of the frameworks significantly depends on the use case implementation. The conclusions that they draw are very similar to their previous study, i.e., Spark is much more efficient than Hadoop but requires higher memory allocation because of its architecture that keeps the data to be processed in node caches.

Ahmed et al. [10] aim to identify the parameters with the highest impact on the performance of Hadoop and Spark by using trial-and-error approach to tune them under a variety of experimental settings. Their evaluation metrics to measure benchmarked frameworks' performance are execution time, throughput, and speedup. Their experimental results show that performance of the frameworks heavily depends on the workload size and correct parameter selection. They show that Spark has better performance when the input size is small, it can get 14 times faster than Hadoop under certain parameters.

Mavridis et al. [11] conduct a log file analysis to highlight the similarities and differences between Spark and Hadoop MapReduce. They use a real very large Apache HTTP Server log file and process it as blocks of 128 MBs in HDFS. They define some metrics to understand the impact of parameters such as input workload size, the number of active nodes, and the type of application. Furthermore, they record utilization information about CPU, memory, disk, and network for each node during experiments.

Liu's research [12] investigates the differences between Spark and MapReduce along with ideal parameters to improve the efficiency. Common applications such as aggregation, shuffle/sort, and iterative jobs are represented by workloads of Wordcount, TeraSort, and K-means, respectively. Their experiments show that Spark performs better than Hadoop for iterative and aggregation jobs, while Hadoop is a better choice for shuffle/sort kinds of jobs.

Lagwankar et al. [13] study micro-benchmarks like PageRank, Grep, WordCount, Sort, Matrix Multiplication, and Fast Fourier Transform (FFT). Clustering applications using Hadoop and Spark are executed under varying dataset sizes and framework combinations. They also examine the effect of underlying frameworks on the behavior. They conclude that these micro-benchmarks' behaviors depend on the size, pattern, type, and source of input datasets.

Vettriselvi et al. [14] compare Spark and Hadoop on HiBench for performance analysis. Varying benchmark types are applied to both frameworks such as TeraSort, WordCount, Logistic Regression, Gradient Boosting Tree, PageRank, etc. Their experimental results show that Spark outperforms Hadoop in terms of execution time and throughput per node.

The above summarized previous benchmark efforts to compare Hadoop MapReduce and Apache Spark generally fall short in terms of specifying evaluation metrics that is specific to the studied task. As an example, the commonly employed execution time metric often refers to the running time of a program, and it does not consider the predominant components of the studied algorithm. We define our evaluation metrics by taking the task-specific factors into consideration. For instance, we refer to execution time to indicate the duration for training the model instead of the execution of the whole program. We also consider an essential metric for classification to measure the prediction success of a trained model, i.e., accuracy. By changing the configuration of the environment setup, such as adding clusters to the environment, we show the speedup of each framework associated with scalability. We also provide composite result tables to show more than one parameter concurrently. Furthermore, workloads that are studied in other works are often the same dataset generated with varying sizes to represent different inputs. Nevertheless, in this work, we work with diverse datasets in varying sizes to imitate real-world scenarios.

## 3 METHODS

We benchmark two popular Big Data Analytics tools, namely, Apache Spark and Hadoop MapReduce, for classification task on Big Data. This section introduces these tools briefly and explains how they are configured to implement a benchmark environment. Then, a representative machine learning algorithm, Naive-Bayes Classiffier, that is used for building a classification model on different datasets, is explained in detail. Finally, our benchmark methodology is presented.

### 3.1 Frameworks and Utilities

#### 3.1.1 Apache Spark

Apache Spark is a Big Data Analytics tool for large-scale data processing. It provides an interface for programming clusters with data parallelism. Spark is able to abstract the application from the data through a concept of resilient distributed dataset (RDD), a read-only multiset of data items distributed over a cluster of machines [15]. We conduct the first part of our benchmarking experiment with Apache Spark on top of Hadoop Distributed File System (HDFS) [16].

We utilize the MLlib library of Spark, which is integrated into Spark with Python language (PySpark) [17]. MLlib exposes some core data science functionalities with PySpark such as Data Engineering Tools, Machine Learning Algorithms, and Utilities.

#### 3.1.2 Hadoop MapReduce

MapReduce is a programming model and an associated implementation for processing big datasets with a parallel and distributed algorithm on clusters [18]. MapReduce program work in two phases, namely, Map and Reduce. Map tasks deal with splitting and mapping of data, while Reduce tasks shuffle and reduce the data.

Hadoop MapReduce is a software framework implementing big data applications on clusters exploiting the MapReduce phenomenon. We employ the Mahout utility on top of Hadoop MapReduce, that offers common machine learning algorithm implementations [19].

### 3.2 Naïve-Bayes Classifier

Naïve-Bayes methods are a set of supervised learning algorithms, reside under linear classifier family's generative models [20]. Bayes' theorem states the following relationship, given class variable y and dependent feature vector $x_1$ through $x_n$:

$$P(y|x1,...,xn) = \frac{P(y)P(x1,...,xn|y)}{P(x1,...,xn)} \quad (1)$$

Using the naive conditional independence assumption that

$$P(xi|y, x1,...,xi-1; xi+1,...,xn) = P(xi|y) \quad (2)$$

for all i, this relationship is reduced to

$$P(y|x1,...,xn) = \frac{P(y) \prod P(xi|y)}{P(x1,...,xn)} \quad (3)$$

Since P(x$_1$, ..., x$_n$) is constant given the input, we can use the following classification rule:

$$y = argmaxP(y) \prod P(xi, y) \quad (4)$$

and here a Maximum A Posteriori (MAP) [21] estimation can be used to estimate P(y) and P(x$_i$ | y); therefore P(y) is the relative frequency of class y in the training set.

Naïve-Bayes Classifier is implemented in each benchmarking process of Big Data Tools by using their machine learning utilities. They are explained in detail for each tool and the source code scripts are provided in our GitLab repository [22].

*3.2.1 MLlib on Spark*

MLlib is Apache Spark's machine learning library [17]. It implements most of the popular machine learning algorithms in a scalable manner. MLlib has a support for Naïve-Bayes Classifier family; Multinomial Naïve-Bayes, Complement Naïve-Bayes, Bernoulli Naïve-Bayes, and Gaussian Naïve-Bayes. We use the Multinomial Naïve-Bayes [20] for our implementation in the benchmark. As application programming interface (API) to take advantage of MLlib utilities, we use PySpark.

*3.2.2 Mahout on MapReduce*

Apache Mahout is a distributed linear algebra framework that implements machine learning algorithms focused primarily on linear algebra [19]. We utilize Mahout's Naïve-Bayes implementation in our experiments to execute the classification tasks on Hadoop MapReduce.

**3.3 Benchmarking Methodology**

Instead of merely sharing the empirical results of the conducted experiments of our study, we point out some necessary parameters and features of the benchmarked frameworks. Besides showing the effect of each configured parameter on performance, we also provide a compound matrix of parameters to show how composite parameter groups affect the performance of the frameworks collectively.

Rather than using existing benchmarking methods such as HiBench, we create a custom benchmarking environment that does not restrict us in terms of defining the metrics to measure the performance. Input workloads of the experiments are different datasets with varying sizes which distinguishes our study than previous benchmark efforts those generate common datasets with various sizes such as TeraGen and WordCount.

Another highlighted feature of our benchmark is that it measures the performance of frameworks in a way that is specific to the studied task, i.e., classification in our case. We define some evaluation metrics in Section IV and apply them in our analysis of the benchmarked frameworks. These specialized metrics help us to comprehend the effect of different configurations on the success and performance of classification.

The Hadoop Distributed File System (HDFS) is used as an infrastructure for distributed file system in the experiment. Apache Hadoop's YARN (Yet Another Resource Negotiator) software is utilized for allocating system resources and scheduling the tasks [23]. Respective core utilities are built on top of YARN to execute data mining operations. The ecosystem of our experimental environment is illustrated in Figure 1.

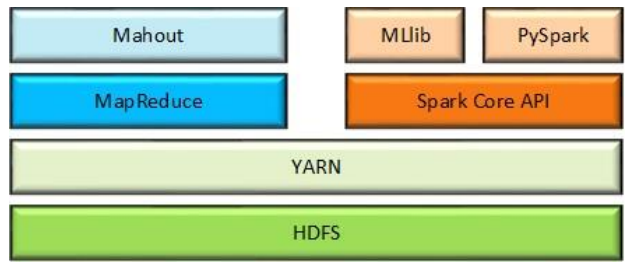

Figure 1: Ecosystem of the experiment

## 4 EVALUATION

### 4.1 Datasets

Most of the prior works generated varying sizes of the same dataset to test frameworks with different sizes of workloads. In this work, we use several different datasets to test our underlying frameworks, all of which are binary classification datasets. These datasets are in LibSVM format [24], for usage of Spark. Then we generate sequential files to be processed by Hadoop utilities. Table 1 shows the employed datasets in our study.

Table 1: Datasets used in experiment

| Dataset | Size |
|---|---|
| HeartBeat | 75 KB |
| URL | 400 MB |
| WebSpam | 2 GB |
| KDD12 | 21 GB |

*4.1.1 HeartBeat Dataset*

HeartBeat dataset from Statlog is a small-size dataset (75 KB) that consists of 270 instances with 13 attributes (age, sex, resting blood pressure, etc.) [25]. The main reason for using such a dataset is to see whether there is an overhead in Big Data tools while processing small workloads.

*4.1.2 URL Reputation Dataset*

It is an anonymized 120-day subset of the ICML-09 URL data [26] containing 2.4 million instances with 3.2 million attributes, which has the size of 400 MB.

*4.1.3 WebbSpam Corpus Dataset*

This dataset contains web pages that are created to deceive Web users and manipulate search engines. It contains 350,000 instances with 16.6 million attributes, and has the size of 2 GB in total [27].

*4.1.4 KDD12 Dataset*

It is a binary classification dataset to predict whether a user is going to follow an item in Tencent Wiebo. It was originally used in KDD Cup 2012, named as Click Through Prediction Competition. The dataset consists of 149.6 million instances with 54.6 attributes, and has a total size of 21 GB [28].

**4.2 Evaluation Metrics**

We define several evaluation metrics to compare the studied frameworks, Apache Spark and Hadoop MapReduce. These metrics are *execution time, accuracy,* and *scalability*. Each metric is defined in terms of measuring the efficiency of the classification task. The way that we employ these metrics to measure the performance is explained in detail below.

*4.2.1 Execution Time*

Execution time in our study refers to the time that it takes to train a model with Naive Bayes Classifier. We use the timer library of Python for Apache Spark. Hadoop MapReduce jobs already measure and log the duration of each task throughout the execution. Note that since we primarily focus on benchmarking the classification task, we only consider the training phase of the model in terms of execution time. Other components of the program such as importing libraries and dataset, the whole running time of the program are not considered in this research, as they were widely studied in earlier works as discussed in Section 2.

*4.2.2 Accuracy*

The accuracy of a data mining algorithm is a way to measure how successful a model is on classifying a data point correctly. We split our datasets into two sets, train and test sets (80% and 20%), and then, employ them for training the classifier model and measure the accuracy, respectively.

*4.2.3 Scalability*

Scalability is a vital dimension since Big Data operations are conducted on distributed systems due to its huge volume. We adjust our cluster configurations to measure the speedup due to adding new processing units.

**4.3 Experimental Setup**

As a project environment, we use Amazon's Elastic MapReduce (EMR) which is an Amazon Web Services (AWS) tool for Big Data processing and analysis [29]. m5.xlarge cluster is configured with 1 master and n slave nodes where n ∈ (2,4,8,16). Datasets are uploaded to the Amazon's Simple Storage Service (Amazon S3) and retrieved from there by EMR clusters. Figure 2 depicts our experimental setup. All the experiments with varying configurations are repeated three times and the mean values of these records are reported.

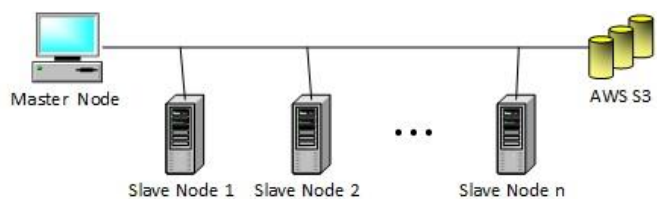

Figure 2: Environment setup of the experiments on AWS

**4.4 Results**

Results are compared for both frameworks, Spark and MapReduce, and discussed in detail in terms of each evaluation metric. We also provide combined tables to show the effect of more than one metric at a time.

*4.4.1 Execution Time*

Experiments show that Spark is much faster than Hadoop during the training phase. As the workload size gets larger, execution time grows linearly for Spark. A similar execution time growth pattern is observed for Hadoop MapReduce, but with much longer time durations. Consistent with the reported results in [8], Spark's performance degrades when the input size gets larger. This is because the way Spark reads the input files and distributes over RDDs which creates features that reside over distributed clusters' memories and dependent to each other during the training phase. MapReduce overcomes this by preparing the vectors from sequential input files during the preprocessing stage, and each cluster exploit these vectors during the training stage. Figure 3 and Table 2 show how input dataset size affects the training time of the classification task.

Table 2: Execution time of the training phase

| Tool / Dataset    | Spark   | MapReduce |
|-------------------|---------|-----------|
| Hearbeat (75 KB)  | 4.48s   | 35.2s     |
| WebbSpam (400 MB) | 8.55s   | 37.6s     |
| URL (2 GB)        | 43.59s  | 40.5s     |
| KDD12 (21 GB)     | 386.6s  | 308.8s    |

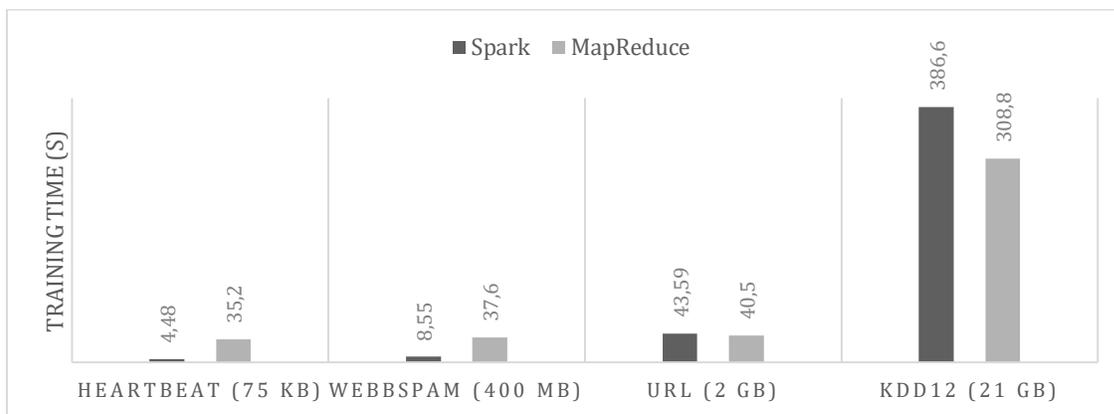

Figure 3: Execution time of the training stage of the frameworks

*4.4.2 Accuracy*

Accuracy scores does not follow a consistent comparison pattern across different datasets. Experiments show that Mahout's Naïve-Bayes Classifier model having better accuracy scores than MLlib. Figure 4 and Table 3 show the change in accuracy for different datasets.

Table 3: Accuracy scores of the frameworks

| Tool / Dataset    | Spark   | MapReduce |
|-------------------|---------|-----------|
| Heartbeat (75 KB) | 69.96%  | 71.8%     |
| WebbSpam (400 MB) | 65.73%  | 66.48%    |
| URL (2 GB)        | 93.5%   | 98.83%    |
| KDD12 (21 GB)     | 98.7%   | 99.5%     |

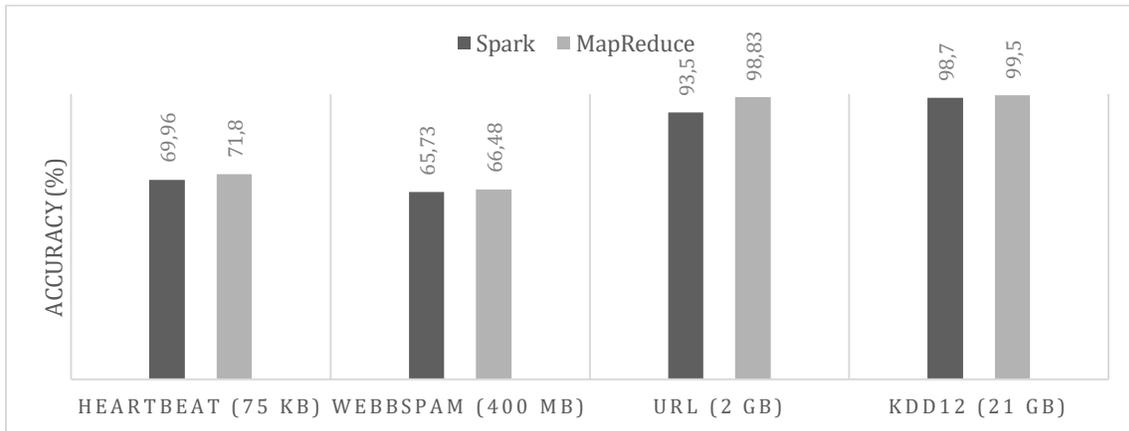

Figure 4: Accuracy scores of frameworks in different datasets

*4.4.3 Scalability*

Increasing the number of nodes is expected to reduce the amount of work on individual computing units. Experiments show that adding more clusters to the system significantly boosts Spark's performance. However, MapReduce does not achieve that much improvement in performance as we add new clusters to the environment. The reason is that Hadoop's operations are input/output bound because of its disk dependency. Adding more clusters to the system means more input/output operations made by extra nodes on the shared disk. Table 4 shows the speedup of each framework with different node and input configurations.

Table 4: Speedup of frameworks with varying number of clusters

| Number of Slave Clusters | Input Size | Spark | MapReduce |
|---|---|---|---|
| 2 | 2 GB | 42s | 41s |
|   | 400 MB | 8s | 42s |
| 4 | 2 GB | 28s | 42s |
|   | 400 MB | 6s | 41s |
| 8 | 2 GB | 19s | 42s |
|   | 400 MB | 4s | 38s |
| 16 | 2 GB | 16s | 43s |
|   | 400 MB | 3s | 39s |

## 5 CONCLUSION AND FUTURE WORK

In this paper, we present a benchmarking study to compare the performance of Apache Spark and Hadoop MapReduce on a common data mining task, i.e. classification. Four different data sets are used to represent varying sizes of input workloads from different domains. The studied frameworks are compared in the aspects of execution time, accuracy, and scalability. Experiments demonstrate that Spark is about 5 times faster than Hadoop in terms of the execution time of training phase. This rate is proportional to the number of clusters in the system, since Spark scales better with increasing number of nodes in the cluster. However, the performance of Spark degrades as the input workload gets larger. Our results show that Hadoop provides consistently better classification accuracy figures than Spark.

As part of our future work, we would like to investigate behavior of the employed systems with larger datasets. Furthermore, other data mining tasks like regression, clustering, etc. could be studied to observe the impact of different data mining tasks on performance.